\documentclass[reprint, preprintnumbers, aps, prl, floatfix, superscriptaddress, nofootinbib]{revtex4-2}

\usepackage{silence}
\usepackage{amssymb, amsmath, physics, graphicx, xcolor, orcidlink, fontawesome, hyperref, bm, mathrsfs}

\newcommand{\DAMTP}{Department of Applied Mathematics and Theoretical Physics, Centre for Mathematical Sciences, University of Cambridge, Wilberforce Road, Cambridge, CB3 0WA, UK}
\newcommand{\SJC}{St John's College, Cambridge, CB2 1TP, United Kingdom}

\begin{document}

\title{Binary black-hole mergers with ultralight vector bosons}

\author{Adrian Ka-Wai Chung\,\orcidlink{0000-0003-2020-3254}}
\email{kwc43@cam.ac.uk}
\affiliation{\SJC} \affiliation{\DAMTP}

\date{\today}

\begin{abstract}
We simulate the inspiral, merger, and ringdown of equal-mass, nonspinning binary black holes coupled to a massive vector (Proca) field. 
Starting from constraint-consistent electric-field initial data, we find that the Proca field rapidly concentrates around the black holes. 
However, it continues to exhibit nontrivial dynamics. Orbital evolution depends on vector mass, while the dominant gravitational waveforms remain close to vacuum after time, phase, and amplitude alignment. 
This contrast highlights the limited imprint of the field on aligned merger-ringdown morphology at these resolutions.
\end{abstract}

\maketitle

{\textit{\textbf{Introduction}}---}Ultralight bosons are candidates for physics beyond the Standard Model and for dark matter \cite{Weinberg_1977, Wilczek_1978, Hui_2017, Ferreira_2021}. 
For sufficiently small masses, their Compton wavelengths can become comparable to the length scales of astrophysical black holes, giving rise to phenomena that can be probed through black-hole and gravitational-wave observations. 
In particular, massive vector fields can undergo superradiant instabilities around rotating black holes, extracting their energy and angular momentum and forming vector-field condensates \cite{Starobinskii:1973vzb,Detweiler_1980,Dolan_2007,Brito_2020,Siemonsen_2020,Baryakhtar_2021,Ng_2021,Tong_2022,Alexander_2022,Fan_2024,May_2025, Pani:2012vp, East:2017ovw}. 
The resulting systems can emit gravitational waves and have motivated searches for ultralight vector bosons with gravitational-wave observations \cite{ Arvanitaki_2011,Arvanitaki_2015,Baryakhtar:2017ngi,Brito:2017wnc,Brito:2017zvb,DAntonio:2018sff,Hannuksela:2018izj,Ghosh:2018gaw,Berti:2019wnn,Tsukada:2018mbp,Tsukada:2020lgt,Ng_2021,Maselli_2021,Duque_2024,Xie_2025,Aswathi:2025nxa,Dyson:2025dlj, Li:2025ffh, Guo:2025pea,Wade_2026,Rahman_2026,Lui_2026c,Li:2026foy}.

Ultralight bosons can also affect binary black holes, altering their gravitational-wave signals. 
The presence of a surrounding field can modify the orbital dynamics, while the tidal field of a companion can perturb or redistribute bosonic configurations around the black holes \cite{Baumann:2018vus, Berti:2019wnn, Siemonsen:2019ebd, Baumann:2021fkf, Tong:2022bbl, Tomaselli:2024dbw}. 
These effects have been studied using perturbative, post-Newtonian, and semi-analytical approaches \cite{Baumann:2018vus, Siemonsen:2019ebd, Chung:2021roh, Hancock:2026ijo}. 
While fully nonlinear numerical-relativity simulations have explored binary black-hole mergers in scalar-field environments \cite{Bamber:2022pbs, Aurrekoetxea:2023jwk, Aurrekoetxea:2024cqd}, corresponding simulations with massive vector fields are still lacking.

In this paper, we perform numerical-relativity simulations of equal-mass, nonspinning binary black holes coupled to a real massive vector field. 
We construct constraint-consistent initial data and evolve the coupled Einstein--Proca system with full nonlinear backreaction. 
For the configurations studied here, we find that the vector field modifies the orbital evolution, while the dominant gravitational-wave morphology remains close to the vacuum result after time, phase, and amplitude alignment. 
Only the numerical formulation and initial-data construction needed for the present analysis are summarized below; further details and numerical checks are provided in the Supplemental Material.

Throughout this work, we use units in which $c=G=\hbar=1$, with all dimensionful quantities expressed in terms of the total binary mass $M$.
We quote $m_V$ and the Proca electric and magnetic fields, including the initial amplitude $E_0$, in units of $M^{-1}$.

{\textit{\textbf{Models of ultralight vector bosons}}---}
We consider a real massive vector field minimally coupled to gravity, 
governed by the following Lagrangian density \cite{Jackson:1998nia, East:2017ovw, Herdeiro:2016tmi}, 
\begin{equation}
\mathscr{L} =
\frac{R}{16\pi}-\frac{1}{4}F_{\mu\nu}F^{\mu\nu}-\frac{1}{2}m_V{}^2 A_\mu A^\mu,
\end{equation}
where
\begin{equation}
F_{\mu\nu}=\nabla_{\mu} A_{\nu} - \nabla_{\nu} A_{\mu}.
\end{equation}
Variation with respect to the metric and the vector field gives the equations of motion
\begin{subequations}
\label{eq:EOM}
\begin{align}
R_{\mu\nu}-\frac{1}{2}g_{\mu\nu}R
&=8\pi T_{\mu\nu},
\label{eq:Einstein}\\
\nabla_\mu F^{\mu\nu}
&=m_V^2 A^\nu.
\label{eq:Proca}
\end{align}
\end{subequations}
where
\begin{equation}
\begin{split}
T_{\mu\nu}
= & 
F_{\mu\rho}F_{\nu}{}^{\rho}
-\frac14 g_{\mu\nu}F_{\rho\sigma}F^{\rho\sigma} \\
& +m_V^2
\left(
A_\mu A_\nu
-\frac12 g_{\mu\nu}A_\rho A^\rho
\right).
\end{split}
\end{equation}
Taking the divergence of Eq.~\eqref{eq:Proca} gives
\begin{equation}
\nabla_\mu A^\mu=0,
\end{equation}
for $m_V\neq0$.
This Lorenz condition follows from the massive vector-field equations
and is not a gauge choice.

{\textit{\textbf{Numerical evolution}}---}The spacetime is evolved using a $W$-BSSN formulation \cite{Baumgarte:1998te, Marronetti:2007wz, Baumgarte:2010ndz, Rezzolla:2013dea} with moving-puncture gauge conditions \cite{Campanelli:2005dd}. 
We evolve the Proca sector by decomposing the vector field into the scalar potential $\Phi=-n^\mu A_\mu$, spatial vector potential $a_i=\gamma_i{}^\mu A_\mu$, and electric field
\begin{equation}
E^i = F^{i\mu}n_\mu,
\end{equation}
while the magnetic field is reconstructed as
\begin{equation}
B^i=\epsilon^{ijk}D_j a_k, 
\end{equation}
where $\epsilon^{ijk}$ is the spatial Levi-Civita tensor associated with $\gamma_{ij}$, normalized such that $\epsilon^{123}=1/\sqrt{\gamma}$, with $\gamma=\det(\gamma_{ij})$.
We use $\alpha$, $\beta^i$, $K$, and $D_i$ to denote the lapse, shift, trace of the extrinsic curvature, and spatial covariant derivative, respectively.
Further details of the $3+1$ decomposition and BSSN formulation are given in the Supplemental Material.
We evolve the variables $(a_i,E^i,\Phi)$ together with an auxiliary constraint-cleaning field $Z$. 
Defining
\begin{equation}\label{eq:Proca_constraint}
\mathcal{G} = D_iE^i+m_V^2\Phi
\end{equation}
as the Proca constraint and
$\partial_\perp\equiv\partial_t-\mathcal{L}_{\boldsymbol{\beta}}$,
the evolution system is
\begin{align}
\partial_\perp a_i &= -\alpha E_i-D_i(\alpha\Phi),
\nonumber\\
\partial_\perp E^i &= \alpha\left( K E^i+m_V^2a^i+D^iZ \right) + \epsilon^{ijk}D_j(\alpha B_k),
\nonumber\\
\partial_\perp\Phi &= \alpha\left( K\Phi-D_ia^i-Z \right) -a^iD_i\alpha,
\nonumber\\
\partial_\perp Z &= \alpha\left( \mathcal{G}-\kappa Z \right).
\label{eq:proca_evolution_main}
\end{align}
The magnetic field satisfies its divergence constraint by construction.

The physical constraint surface is $\mathcal{G}=Z=0$. 
The scalar-potential evolution equation implies the generalized Lorenz relation
\begin{equation}
\nabla_\mu A^\mu=-Z,
\end{equation}
and therefore recovers the Lorenz condition $\nabla_\mu A^\mu=0$ on the physical constraint surface. 
The auxiliary field also converts violations of the Gauss constraint into a propagating cleaning subsystem, 
\begin{equation}
\partial_\perp \mathcal{G}
=
\alpha K \mathcal{G}
+
D_i\left(\alpha D^iZ\right)
-
\alpha m_V^2 Z,
\end{equation}
which, together with Eq.~\eqref{eq:proca_evolution_main}, reduces in the Minkowski limit to
\begin{equation}
\partial_t^2Z-\nabla^2Z
+\kappa\partial_tZ
+m_V^2Z
=
0.
\label{eq:cleaning_wave_main}
\end{equation}
Thus numerical violations of the Proca constraint propagate through the cleaning system and are damped for $\kappa>0$; we use $\kappa M=1$. The field $Z$ is auxiliary and does not contribute to the physical Proca stress-energy tensor.

The Proca stress-energy tensor is included in the BSSN evolution equations, so that the vector field and spacetime evolve with full nonlinear backreaction. 
The equations are implemented using NRPy code generation \cite{Ruchlin:2017com} within the Dendro-GR framework \cite{Fernando:2018mov}, with Cartesian octree adaptive mesh refinement, fourth-order Runge--Kutta time integration, sixth-order centered finite differences, and Kreiss--Oliger dissipation \cite{kreiss1973methods}. 
No spatial symmetry is imposed. 
The computational domain extends to $2048M$ in each Cartesian direction, with Courant factor $0.25$ and maximum refinement depth $17$. 

{\textit{\textbf{Initial data}}---} We model a binary black hole embedded in a massive-vector field that is initially approximately uniform across the binary. 
Specifically, we consider two equal-mass, nonspinning punctures with separation $d \sim 12.2M$. 
The initial separation and momenta of the punctures are chosen following~\cite{Andrade2021}, such that the black holes complete approximately $10.25$ orbits before merger.

For the initial Proca field, we choose $a_i=\Phi=B^i=0$, while $E^i$ is given by
\begin{equation}
E^i=\psi^{-6}\hat E^i,
\end{equation}
with
\begin{equation}
\hat E^i
=
E_0e^{-r^2/R^2}
\left[
\left(1-\frac{r^2}{R^2}\right)p^i
+
\frac{(\boldsymbol p\cdot\boldsymbol r)r^i}{R^2}
\right],
\label{eq:main_Eseed}
\end{equation}
where $R=80M$, $E_0=5\times10^{-4}M^{-1}$, and $\boldsymbol p=\boldsymbol e_z$. 
For these parameters, the weak-field estimate of the total Proca-field energy is approximately $0.08M$, or about $8\%$ of the binary mass scale.
Since $\partial_i\hat E^i=0$, the conformal scaling above gives $D_iE^i=0$; together with $\Phi=0$, the Proca constraint is therefore satisfied initially.
This form of $\hat E^i$ is chosen such that the initial field is approximately uniform across the binary while decaying at large radius, since a homogeneous field with nonzero energy density cannot extend to spatial infinity while preserving both asymptotic flatness and finite total energy.

Since $a_i=\Phi=B^i=0$, the matter momentum density also vanishes, $S_i=0$. 
The standard Bowen--York solution of the momentum constraint can therefore be retained, while the Hamiltonian constraint is solved including the Proca energy density to determine $\psi$. 
The initial geometry is thus not inherited from a vacuum binary, but is obtained from a matter-constraint-solved slice that already contains the gravitational backreaction of the Proca field. 
The resulting initial configuration is not a stationary equilibrium configuration. 

\begin{figure*}[t!]
    \centering
    \includegraphics[width=\linewidth]{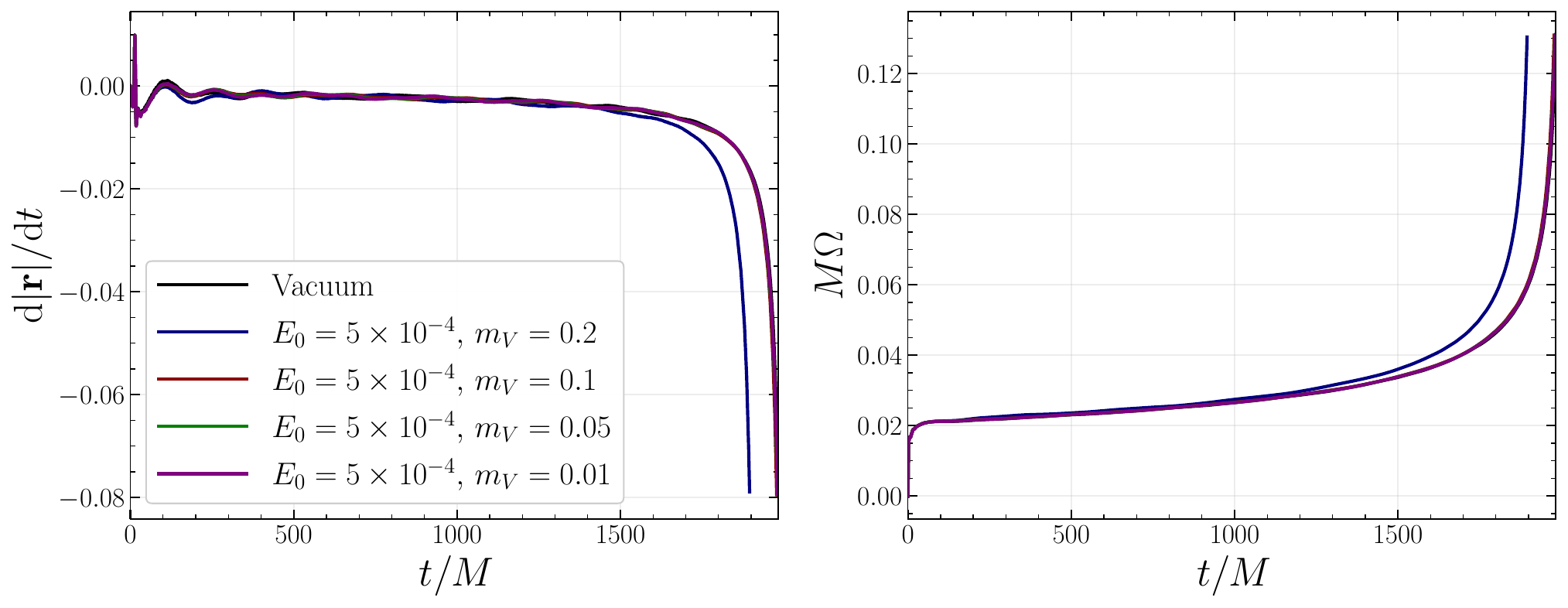}
    \caption{
    Evolution of the radial separation rate $\text{d}|\textbf{r}|/\text{d}t$ (left) and orbital angular frequency $M\Omega$ (right) of the binary black holes for the vacuum and Proca-field simulations with $E_0=5\times10^{-4}M^{-1}$. Both quantities are computed from the coordinate trajectories of the black-hole punctures. The curves are truncated when the coordinate separation first reaches $3M$, the threshold that activates the search for a common apparent horizon. 
    }
    \label{fig:orbital_dynamics}
\end{figure*}

{\textit{\textbf{Orbital dynamics}}---}The presence of the Proca field affects the orbital evolution of the binary black holes. 
Figure~\ref{fig:orbital_dynamics} shows the evolution of the radial separation rate $\text{d}|\textbf{r}|/\text{d}t$, where $\textbf{r}$ is the coordinate separation vector between the two punctures, and the orbital angular frequency $\Omega$.
For $m_V=0.2$, the radial separation decreases more rapidly during the late inspiral, accompanied by a faster increase in the orbital angular frequency, leading to an earlier merger compared to the vacuum case. 
For smaller vector masses, the orbital evolution remains closer to the vacuum result. 

Since the initial Proca stress-energy and geometry are independent of $m_V$ at fixed $E_0$ for the electric-phase initial data considered here, this dependence on $m_V$ arises dynamically during the evolution.
One possible explanation is the massive dispersion relation, $\omega^2=k^2+m_V^2$, which implies a smaller group velocity for larger $m_V$ at fixed spatial wavenumber, potentially allowing a larger fraction of the Proca energy to remain in the near zone and thereby enhancing its backreaction on the binary.
In addition, the threshold between propagating and evanescent modes occurs at $\omega=m_V$; increasing $m_V$ therefore raises this threshold, so that a larger fraction of the Proca-field spectrum may become sub-threshold and therefore less efficient at propagating away from the binary.
Both effects could increase the amount of Proca energy retained in the near zone for larger $m_V$.
Establishing this mechanism, however, requires direct diagnostics of the Proca energy distribution and flux, which we plan to investigate in future work.
These results indicate a dependence of the orbital dynamics on the vector mass at the resolution considered.
However, the quantitative differences in merger time exhibit substantial resolution dependence, as discussed in the Supplemental Material.

{\textit{\textbf{Scaled and aligned waveforms}}---}We extract the fourth Weyl scalar, $\Psi_4$, at several coordinate radii between $50M$ and $110M$. Throughout this work, we use $\Psi_4$ to characterize the emitted gravitational radiation. 
We focus on the waveform extracted at $r=110M$, the largest extraction radius considered, where finite-radius effects are expected to be smallest. 
The propagation of the waveform across different extraction radii is examined in the Supplemental Material as a finite-radius consistency check.

\begin{figure}[htp!]
    \centering
    \includegraphics[width=\linewidth]{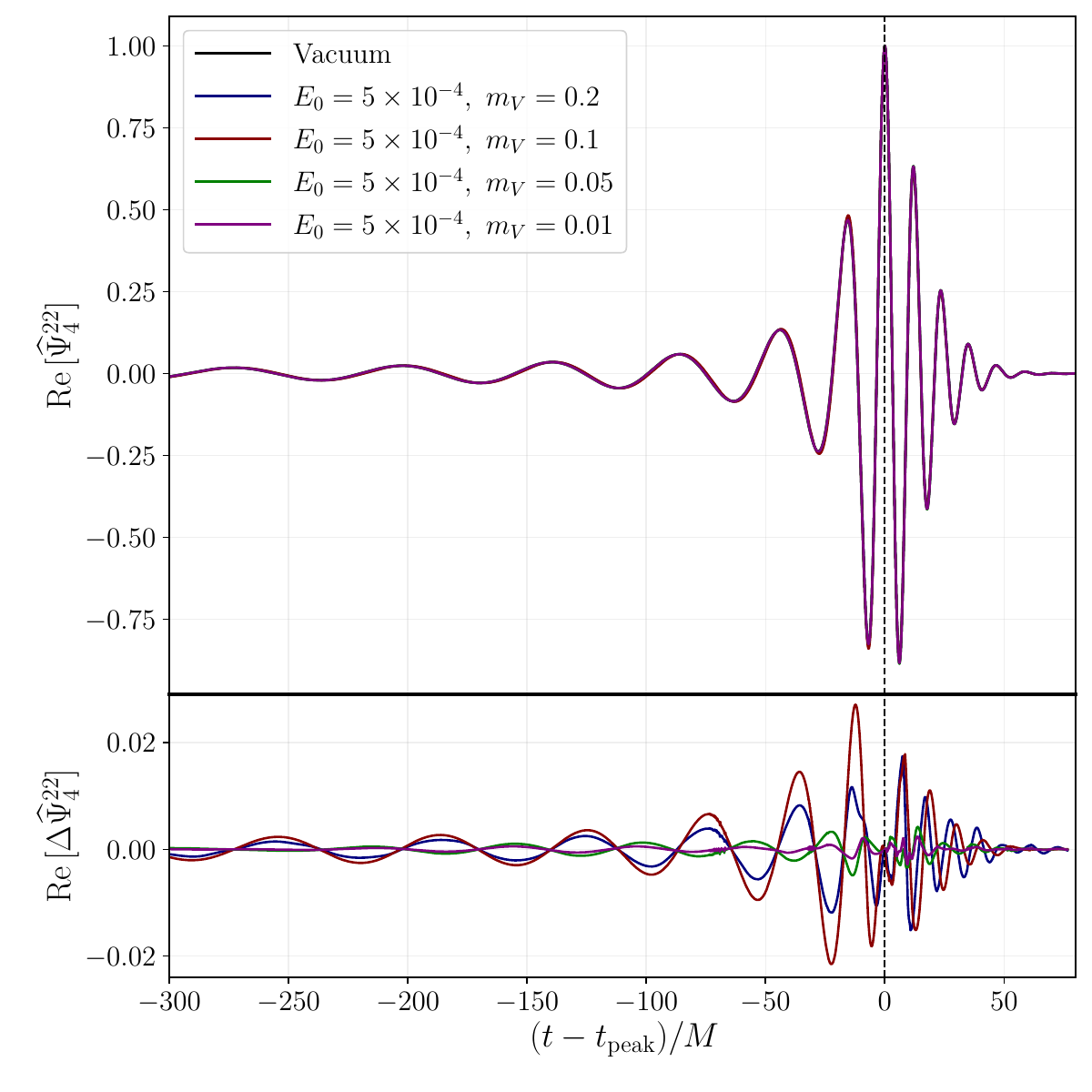}
    \caption{
    Scaled and aligned $(\ell,m)=(2,2)$ gravitational waveforms for the vacuum and Proca-field simulations with $E_0=5\times10^{-4}$. 
    In the upper panel, the waveforms are normalized by their respective peak amplitudes and aligned such that the peak occurs at $t-t_{\rm peak}=0$, with the corresponding phases aligned at the peak. 
    The lower panel shows the residuals relative to the vacuum waveform, $\Delta\widehat{\Psi}_4^{22}\equiv \widehat{\Psi}_{4,\mathrm{Proca}}^{22}-\widehat{\Psi}_{4,\mathrm{vac}}^{22}$. 
    The vertical dashed line marks the peak time.
    }
    \label{fig:Scaled_Psi4}
\end{figure}

\begin{figure*}[t!]
    \centering
    \includegraphics[width=\linewidth]{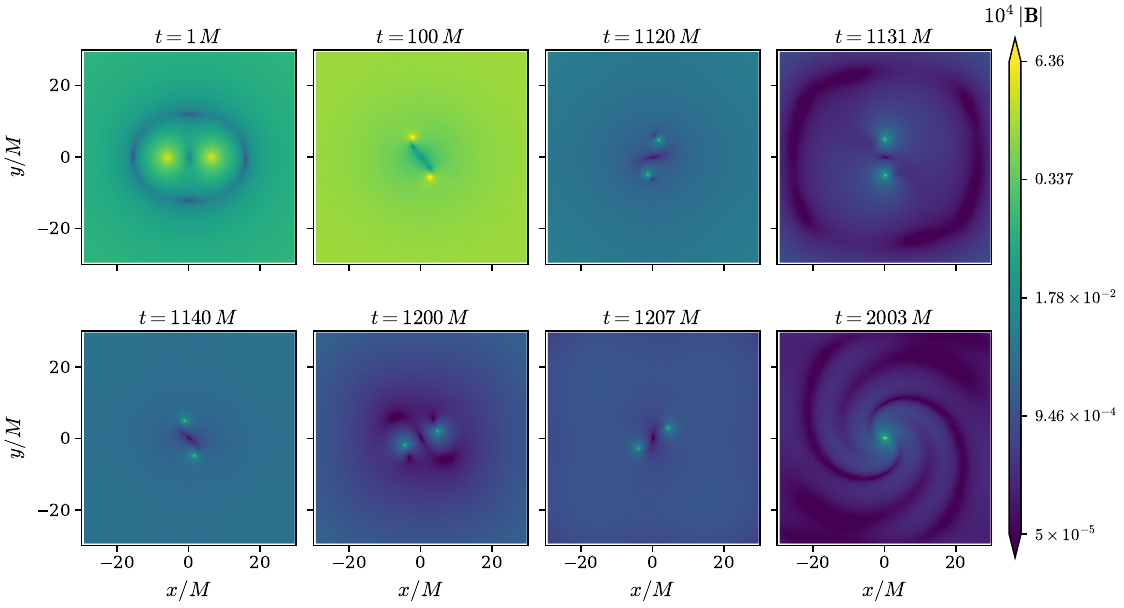}
    \caption{Evolution of the Proca magnetic-field magnitude $10^4|\mathbf{B}|$ in the orbital plane for the $m_V=0.01$ binary-black-hole simulation at different times.
    Note that $|\textbf{B}|=0$ at $t=0$. 
    A common logarithmic color scale is used for all panels.
    }
    \label{fig:field_evolution}
\end{figure*}

The upper panel of Fig.~\ref{fig:Scaled_Psi4} shows the scaled and aligned $(\ell,m)=(2,2)$ mode of $\Psi_4$ obtained for different values of $m_V$ and $E_0$.
To isolate differences in waveform morphology, we remove an overall amplitude, time translation, and constant phase rotation. 
For each waveform, we define $t_{\rm peak}$ as the time at which
$|\Psi_4|$ reaches its maximum and introduce the shifted time $\tau=t-t_{\rm peak}$. 
We then define the normalized waveform
\begin{equation}
\hat{\Psi}_4(\tau)
=
\frac{
\Psi_4(\tau+t_{\rm peak})
}{
\Psi_4(t_{\rm peak})
},
\label{eq:scaled_aligned_psi4}
\end{equation}
such that $\hat{\Psi}_4(0)=1$. 
The normalization removes the overall waveform amplitude, which is degenerate with extrinsic parameters such as luminosity distance, while the time and phase alignments remove the corresponding constant shifts. 
This allows us to compare directly the intrinsic time dependence of the waveforms around merger and ringdown.
As seen in the upper panel, all scaled-and-aligned $\widehat{\Psi}^{22}_4$ waveforms overlap closely, indicating that the effect of the Proca field on the waveform morphology is small after the differences in peak time, phase, and amplitude have been removed. 
This feature is surprising, given that the amplitude of the initial vector-field configuration is chosen such that the initial total Proca-field energy is approximately $0.08M$.

To further investigate the differences between $\widehat{\Psi}^{22}_4$, the lower panel of Fig.~\ref{fig:Scaled_Psi4} shows the residuals between the vacuum waveform and the corresponding waveforms evolved in the presence of a Proca field. 
Around $\tau=0$, the residuals develop an oscillatory, damped structure that persists into the post-merger regime. 
This behavior indicates a difference in the post-merger oscillatory content relative to the vacuum case and is consistent with a change in the ringdown signal. 
At this stage, however, we do not attribute the effect to a modification of the quasinormal-mode spectrum itself. 
Such a difference may arise from changes in the remnant mass or spin, from the surrounding Proca field, or from a combination of these effects. 
A dedicated ringdown analysis is required to disentangle these possibilities.

{\textit{\textbf{Evolution of vector field}}---}To understand the origin of this mild dependence of $\widehat{\Psi}^{22}_4$ on $m_V$, we examine the evolution of the Proca sector. 

To illustrate the evolution of the Proca sector, we focus on the Proca magnetic field.
The evolution of the Proca magnetic field is shown in Fig.~\ref{fig:field_evolution} for the $m_V=0.01$ simulation. 
Although $\mathbf{B}=\mathbf{0}$ at $t=0$, the Proca magnetic field rapidly develops concentrations around the two punctures within the first $M$ of evolution. 
This behavior provides a possible explanation for the mild dependence of $\widehat{\Psi}^{22}_4$ on $m_V$. 
As a substantial fraction of the Proca-field energy becomes concentrated in the vicinity of the black holes, rather than remaining broadly distributed around the binary, the orbital dynamics may remain close to that of a vacuum equal-mass nonspinning binary after the overall mass scale is accounted for. 
This may explain why the scaled-and-aligned waveform remains close to the vacuum waveform.

By $t=100M$, the magnetic field has developed a configuration associated with the binary and remains concentrated around the two punctures. 
During the subsequent inspiral, the magnetic-field distribution undergoes repeated changes in morphology, accompanied by bursts of magnetic-field activity. 
The snapshots between $t=1120M$ and $1207M$ illustrate one such episode, during which the orientation and structure of the magnetic-field lobes are reorganized over the interval. 
Similar rearrangements occur repeatedly throughout the coalescence. 
Near merger, as represented by the $t=2003M$ snapshot, the magnetic field develops a multipolar structure around the binary.

After merger, the remnant black hole settles toward a stationary state through the emission of ringdown gravitational waves characterized by its quasinormal modes [cf.~Fig.~\ref{fig:Scaled_Psi4}]. 
Correspondingly, the Proca field settles into an approximately axisymmetric configuration that continues to oscillate with a characteristic period of $\mathcal{O}(10^2)M$.

{\textit{\textbf{Concluding remarks}}---}
We have developed a fully nonlinear numerical-relativity framework for simulating binary black-hole mergers coupled to a massive vector field, together with constraint-consistent initial data. 
For the configurations considered here, the Proca field modifies the orbital dynamics, while the dominant gravitational-wave morphology remains close to the vacuum result after time, phase, and amplitude alignment.

Our formalism opens several directions for further study. 
The simulations presented here consider a localized electric-phase Proca configuration that is approximately uniform across the binary but is not an equilibrium state. 
Exploring other initial vector-field configurations, particularly configurations that remain concentrated around the binary for longer times, may lead to stronger effects on the binary dynamics and gravitational-wave emission. 
The evolution scheme can also be implemented in other numerical-relativity codes (e.g. \cite{Andrade2021, Pretorius:2004jg, Lindblom:2005qh, spectrecode, Szilagyi:2009qz}) to enable independent validation and to explore alternative initial configurations and numerical formulations. 
More broadly, the same framework provides a starting point for numerical studies of black holes with vector hair \cite{Fan:2016jnz, Herdeiro:2016tmi}. 
These developments will help establish waveform models for gravitational-wave searches for ultralight vector bosons.

{\textit{\textbf{Acknowledgments}}---}This work employs large language models (LLMs), ChatGPT 5.6 Sol Max for manuscript structuring and drafting. 
The author maintained strict oversight over all LLM applications and rigorously reviewed all generated content.  
A.K.W.C. thanks Joan Núñez Corbacho, Daniela Cros, Richard Dyer, Tomasz Kinowski, Carolin Kohl, Leif Lui, Gareth Marks, Christopher Moore, Ulrich Sperhake, and Seppe Staelens for helpful discussions.
A.K.W.C. acknowledges full support from the Title-E Fellowship at St John's College, Cambridge, and partial support from the Herchel Smith Fellowship at the University of Cambridge.
The calculations and results reported in this paper were produced using the computational resources of the Illinois Campus Cluster, which is a computing resource that is operated by the Illinois Campus Cluster Program (ICCP) in conjunction with National Center for Supercomputing Applications (NCSA).

{\textit{\textbf{Data Availability}}---}The numerical-relativity code and associated analysis tools used in this work are publicly available at \url{https://gitlab.com/akwchung2026/gr-proca-bbh}. 
The numerical data underlying the results presented in this work, together with animations of the simulations, are available on Zenodo at DOI: \url{10.5281/zenodo.24248686}. 

\vspace{1cm}
\bibliographystyle{apsrev4-1}
\bibliography{references}

\clearpage
\onecolumngrid
\newpage
\begin{center}
  \textbf{\large{Supplemental Material}} \\
\end{center}
\twocolumngrid

\setcounter{equation}{0}
\setcounter{figure}{0}
\setcounter{table}{0}

\renewcommand{\theequation}{S\arabic{equation}}
\renewcommand{\thefigure}{S\arabic{figure}}
\renewcommand{\thetable}{S\arabic{table}}

\section{3+1 and BSSN formulation}

\subsection{Gravitational sectors}

We decompose the spacetime metric as
\begin{equation}
ds^2
=
-\alpha^2 dt^2
+
\gamma_{ij}
\left(dx^i+\beta^i dt\right)
\left(dx^j+\beta^j dt\right),
\end{equation}
where $\alpha$ is the lapse function, $\beta^i$ is the shift vector,
and $\gamma_{ij}$ is the spatial metric. The extrinsic curvature is
defined by
\begin{equation}
K_{ij}
=
-\frac{1}{2\alpha}
\left(
\partial_t-\mathcal{L}_{\boldsymbol{\beta}}
\right)
\gamma_{ij}.
\end{equation}

We evolve the spacetime using the BSSN formulation with the conformal
factor
\begin{equation}
W=e^{-2\phi},
\end{equation}
such that
\begin{equation}
\gamma_{ij}
=
W^{-2}\tilde{\gamma}_{ij},
\qquad
\det\tilde{\gamma}_{ij}=1.
\end{equation}
The trace and conformal trace-free parts of the extrinsic curvature are
defined by
\begin{equation}
K
=
\gamma^{ij}K_{ij},
\end{equation}
and
\begin{equation}
\tilde{A}_{ij}
=
W^2
\left(
K_{ij}
-\frac{1}{3}\gamma_{ij}K
\right),
\qquad
\tilde{\gamma}^{ij}\tilde{A}_{ij}=0.
\end{equation}

The matter source terms entering the BSSN equations are obtained from
the projections of the stress-energy tensor,
\begin{align}
\rho
&=
n^\mu n^\nu T_{\mu\nu},
\\
S_i
&=
-\gamma_i{}^\mu n^\nu T_{\mu\nu},
\\
S_{ij}
&=
\gamma_i{}^\mu\gamma_j{}^\nu T_{\mu\nu},
\\
S
&=
\gamma^{ij}S_{ij},
\end{align}
where $n^\mu$ is the future-directed unit normal to the spatial
hypersurfaces.

Defining the advective time derivative
\begin{equation}
\partial_0
\equiv
\partial_t-\beta^k\partial_k,
\end{equation}
the BSSN evolution equations take the form
\begin{align}
\partial_0 W
&=
\frac{1}{3}W
\left(
\alpha K-\partial_k\beta^k
\right),
\\
\partial_0\tilde{\gamma}_{ij}
&=
-2\alpha\tilde{A}_{ij}
+
2\tilde{\gamma}_{k(i}\partial_{j)}\beta^k
-
\frac{2}{3}\tilde{\gamma}_{ij}\partial_k\beta^k,
\\
\partial_0 K
&=
-D^iD_i\alpha
+
\alpha
\left(
\tilde{A}_{ij}\tilde{A}^{ij}
+\frac{1}{3}K^2
\right)
\nonumber \\
& \quad +
4\pi\alpha(\rho+S),
\\
\partial_0\tilde{A}_{ij}
&=
W^2
\left[
-D_iD_j\alpha
+
\alpha
\left(
{}^{(3)}R_{ij}
-
8\pi S_{ij}
\right)
\right]^{\rm TF}
\nonumber\\
&\quad
+
\alpha
\left(
K\tilde{A}_{ij}
-
2\tilde{A}_{ik}\tilde{A}^{k}{}_{j}
\right)
+
2\tilde{A}_{k(i}\partial_{j)}\beta^k 
\nonumber \\
& \quad -
\frac{2}{3}\tilde{A}_{ij}\partial_k\beta^k.
\end{align}
Here $D_i$ denotes the covariant derivative associated with
$\gamma_{ij}$, ${}^{(3)}R_{ij}$ is its Ricci tensor, and ${\rm TF}$
denotes the trace-free part with respect to the spatial metric.

For a Cartesian reference metric, the conformal-connection evolution
can be written in terms of the contracted conformal connection
$\tilde{\Gamma}^i$ as
\begin{equation}
\begin{split}
& \partial_0\tilde{\Gamma}^i\\
&=-2\tilde{A}^{ij}\partial_j\alpha
\\
& \quad +
2\alpha
\left(
\tilde{\Gamma}^{i}{}_{jk}\tilde{A}^{jk}
-\frac{2}{3}\tilde{\gamma}^{ij}\partial_j K
-\frac{3}{W}\tilde{A}^{ij}\partial_j W
-8\pi\tilde{\gamma}^{ij}S_j
\right)\\
&\quad
+
\frac{2}{3}\tilde{\Gamma}^i\partial_j\beta^j
+
\frac{1}{3}\tilde{\gamma}^{li}
\partial_l\partial_j\beta^j
+
\tilde{\gamma}^{lj}
\partial_l\partial_j\beta^i
-
\tilde{\Gamma}^j\partial_j\beta^i .
\end{split}
\end{equation}
In the numerical implementation, the conformal connection is evolved
using the reference-metric variable $\bar{\Lambda}^i$ employed by NRPy/Dendro-GR, which reduces to the above Cartesian conformal-connection
form for the flat reference metric used here.

The corresponding Hamiltonian and momentum constraints are
\begin{align}
\mathcal{H}
&=
{}^{(3)}R
+
K^2
-
K_{ij}K^{ij}
-
16\pi\rho
=
0,
\\
\mathcal{M}_i
&=
D_j
\left(
K^j{}_i-\delta^j{}_i K
\right)
-
8\pi S_i
=
0.
\end{align}

\subsection{Proca fields}

We decompose the Proca potential with respect to the timelike unit
normal $n^\mu$ as
\begin{equation}
A_\mu
=
a_\mu+\Phi n_\mu,
\qquad
\Phi=-n^\mu A_\mu,
\qquad
a_\mu n^\mu=0.
\end{equation}
The electric and magnetic fields measured by Eulerian observers are
defined by
\begin{equation}
E^i
=
F^{i\nu}n_\nu,
\qquad
B^i
=
\epsilon^{ijk}D_j a_k,
\end{equation}
where $D_i$ is the covariant derivative associated with the spatial
metric $\gamma_{ij}$ and $\epsilon^{ijk}$ is the spatial Levi-Civita
tensor.

In terms of these variables, the projections of the Proca
stress-energy tensor are
\begin{equation}
\rho =\frac{1}{2}\left[E^2+B^2+m_V^2\left(\Phi^2+a^2\right)\right],
\end{equation}
where $E^2\equiv E_iE^i=\gamma_{ij}E^iE^j$, $B^2\equiv B_iB^i=\gamma_{ij}B^iB^j$, $a^2\equiv a_i a^i=\gamma^{ij}a_i a_j$, and
\begin{equation}
\begin{split}
S_i&=\epsilon_{ijk}E^jB^k+m_V^2\Phi a_i, \\
S_{ij} & = -E_iE_j-B_iB_j+\frac{1}{2}\gamma_{ij}\left(E^2+B^2\right) \\
& \quad + m_V^2\left[a_i a_j +\frac{1}{2}\gamma_{ij}\left(\Phi^2-a^2\right)\right].
\end{split}
\end{equation}
Their spatial trace is
\begin{equation}
S=\frac{1}{2}\left(E^2+B^2\right)+\frac{1}{2}m_V^2\left(3\Phi^2-a^2\right).
\end{equation}

For the numerical evolution we introduce an auxiliary constraint-cleaning
field $Z$ and evolve the eight real variables
\begin{equation}
\left(a_i,E^i,\Phi,Z\right),
\end{equation}
while reconstructing $B^i$ from the curl of $a_i$.
Defining
\begin{equation}
\partial_\perp
\equiv
\partial_t-\mathcal{L}_{\boldsymbol{\beta}},
\end{equation}
the evolution system is
\begin{align}
\partial_\perp a_i
&=
-\alpha E_i
-
D_i(\alpha\Phi),
\\
\partial_\perp E^i
&=
\alpha
\left(
K E^i
+
m_V^2 a^i
+
D^i Z
\right)
+
\epsilon^{ijk}D_j(\alpha B_k),
\\
\partial_\perp\Phi
&=
\alpha
\left(
K\Phi
-
D_i a^i
-
Z
\right)
-
a^iD_i\alpha,
\\
\partial_\perp Z
&=
\alpha
\left(
\mathcal{G}-\kappa Z
\right),
\label{eq:Z_cleaning}
\end{align}
where $\kappa>0$ controls the damping of the constraint-cleaning
subsystem.

The Lorenz condition is not an independent gauge condition in the
massive Proca theory. Instead, it follows from the Proca equations of
motion. In the $3+1$ decomposition,
\begin{equation}
\mathcal{L}_A
\equiv
\nabla_\mu A^\mu
=
\frac{1}{\alpha}\partial_\perp\Phi
-
K\Phi
+
D_i a^i
+
a^iD_i\ln\alpha .
\end{equation}
The cleaned evolution equation for $\Phi$ therefore implies
\begin{equation}
\mathcal{L}_A+Z=0.
\end{equation}
On the physical constraint surface,
\begin{equation}
\mathcal{G}=Z=0,
\end{equation}
and hence the Lorenz relation
\begin{equation}
\nabla_\mu A^\mu=0
\end{equation}
is recovered.

The cleaning variables obey Eq.~\eqref{eq:Z_cleaning} and 
\begin{align}
\partial_\perp \mathcal{G}
&=
\alpha K \mathcal{G}
+
D_i\left(\alpha D^i Z\right)
-
\alpha m_V^2 Z.
\end{align}
In the Minkowski limit, these equations reduce to
\begin{equation}
\partial_t^2 Z
-
\nabla^2 Z
+
\kappa\,\partial_t Z
+
m_V^2 Z
=
0,
\end{equation}
showing that violations of the $(\mathcal{G},Z)$ constraint subsystem
propagate and are damped for $\kappa>0$.

\subsection{Lapse and shift gauge conditions}

We adopt the moving-puncture gauge conditions used in the underlying NRPy/Dendro-GR evolution. The lapse is evolved using the advective $1+\log$ slicing condition \cite{Bona:1994dr, Anninos:1995am},
\begin{equation}
\left(
\partial_t-\beta^i\partial_i
\right)\alpha
=
-2\alpha K.
\end{equation}
For the puncture initial data, the lapse is initialized according to
\begin{equation}
\alpha(t=0)=W=\psi^{-2}.
\end{equation}

The shift is evolved using a non-advecting Gamma-driver condition.
To avoid confusion with the Proca magnetic field $B^i$, we denote
the auxiliary shift-driver field by $B_{\rm sh}^i$. The gauge
equations are
\begin{align}
\partial_t\beta^i
&=
B_{\rm sh}^i,
\\
\partial_t B_{\rm sh}^i
&=
\frac{3}{4}\,
\partial_t\bar{\Lambda}^i
-
\eta B_{\rm sh}^i,
\end{align}
where $\bar{\Lambda}^i$ denotes the evolved reference-metric
conformal-connection variable and $\eta$ is the shift-damping
parameter. We use
\begin{equation}
\eta M = 1.
\end{equation}
No explicit shift-advection terms are included in the Gamma-driver
sector.

In the numerical implementation, the driver is constructed from the pre-dissipation right-hand side of $\bar{\Lambda}^i$. 
Consequently, the matter contribution to the conformal-connection equation is also included consistently in the driver equation.

\section{Initial data}
\label{sec:initial_data}

We construct asymptotically flat puncture initial data describing two
nonspinning black holes immersed in a localized real Proca field.
We adopt a conformally flat and maximally sliced spatial geometry,
\begin{equation}
\gamma_{ij}
=
\psi^4 \delta_{ij},
\qquad
K=0,
\end{equation}
and decompose the extrinsic curvature according to
\begin{equation}
K_{ij}
=
\psi^{-2}\hat{A}_{ij}.
\end{equation}
The conformal factor is written in puncture form as
\begin{equation}
\psi
=
1
+
\frac{m_1}{2r_1}
+
\frac{m_2}{2r_2}
+
u,
\label{eq:puncture_psi}
\end{equation}
where $m_A$ and $r_A$ denote the bare mass parameter and coordinate
distance from puncture $A$, respectively, and $u$ is regular at the
punctures.

For nonspinning punctures, we take the conformal extrinsic curvature
to be the standard Bowen--York solution \cite{Bowen:1980yu},
\begin{equation}
\hat{A}^{ij}
=
\sum_A
\frac{3}{2r_A^2}
\left[
P_A^i n_A^j
+
P_A^j n_A^i
-
\left(
\delta^{ij}
-
n_A^i n_A^j
\right)
P_A^k n^A_k
\right],
\label{eq:BY_momentum}
\end{equation}
where
\begin{equation}
n_A^i
=
\frac{x^i-x_A^i}{r_A}.
\end{equation}

\subsection{Solenoidal electric-phase Proca configuration}

We choose the initial Proca field at an electric phase,
\begin{equation}
a_i=0,
\qquad
\Phi=0,
\qquad
B^i=0.
\label{eq:candidate1_phase}
\end{equation}
The physical electric field is prescribed through a conformal field
$\hat{E}^i$,
\begin{equation}
E^i
=
\psi^{-6}\hat{E}^i.
\label{eq:E_conformal_scaling}
\end{equation}

To construct a smooth divergence-free electric field, we introduce
an auxiliary vector
\begin{equation}
\boldsymbol{\mathcal C}
=
\frac{E_0}{2}
e^{-r^2/R^2}
\left(
\boldsymbol{p}\times\boldsymbol{r}
\right),
\end{equation}
where
\begin{equation}
\boldsymbol{r}
=
\boldsymbol{x}-\boldsymbol{x}_c,
\qquad
\boldsymbol{p}\cdot\boldsymbol{p}=1.
\end{equation}
We then define
\begin{equation}
\hat{\boldsymbol E}
=
\nabla\times\boldsymbol{\mathcal C}.
\end{equation}
Explicitly,
\begin{equation}
\hat{E}^i
=
E_0 e^{-r^2/R^2}
\left[
\left(
1-\frac{r^2}{R^2}
\right)p^i
+
\frac{
(\boldsymbol{p}\cdot\boldsymbol{r})r^i
}{R^2}
\right].
\label{eq:solenoidal_E}
\end{equation}
By construction,
\begin{equation}
\partial_i\hat{E}^i=0.
\end{equation}

With the conformal scaling in Eq.~\eqref{eq:E_conformal_scaling},
the physical divergence satisfies
\begin{equation}
D_i E^i
=
\psi^{-6}\partial_i\hat{E}^i
=
0.
\end{equation}
Together with $\Phi=0$, the Proca constraint therefore
vanishes identically on the initial slice,
\begin{equation}
\mathcal{G}=0.
\end{equation}
Furthermore, since $a_i=\Phi=B^i=0$, the matter momentum density
vanishes, $S_i = 0$.
Consequently, the gravitational momentum constraint is unchanged from the vacuum Bowen--York problem, and no additional vector elliptic equation is required.

The energy density on the initial slice reduces to
\begin{equation}
\rho
=
\frac{1}{2}E_iE^i
=
\frac{1}{2}\psi^{-8}\hat{E}^2.
\end{equation}
The Hamiltonian constraint therefore becomes a single nonlinear
elliptic equation for the regular correction $u$,
\begin{equation}
\nabla^2 u
+
\frac{
\hat{A}_{ij}\hat{A}^{ij}
}{
8\psi^7
}
+
\frac{
\pi\hat{E}_i\hat{E}^i
}{
\psi^3
}
=
0,
\label{eq:proca_hamiltonian}
\end{equation}
where $\nabla^2$ denotes the flat-space Laplacian associated with the
conformal metric $\delta_{ij}$.

The construction represents a localized coherent vector field at an
electric phase. For $R$ much larger than the binary separation, the
field is approximately uniform over the binary region, while its
Gaussian envelope ensures finite total matter energy and
asymptotically flat boundary conditions. The configuration should
therefore be interpreted as a finite localized Proca environment
rather than a globally homogeneous cosmological background.

\section{Outer boundary conditions}
\label{sec:outer_boundary}

We treat the gravitational and Proca outer-boundary conditions
separately. For the gravitational variables, we retain the
Sommerfeld-type outer-boundary treatment of the underlying
NRPy/Dendro-GR BSSN implementation.

For the Proca sector, a massless Sommerfeld condition is not an exact
absorbing boundary condition. In the asymptotic weak-field limit,
physical Proca modes satisfy
\begin{equation}
\omega^2
=
k^2+m_V^2,
\end{equation}
and therefore have a frequency-dependent group velocity
\begin{equation}
v_g
=
\frac{\partial\omega}{\partial k}
=
\frac{k}{\omega}.
\end{equation}
We instead impose a leading-order characteristic boundary condition
on the incoming Proca modes.

At a distant outer boundary, let $s^i$ denote the outward-pointing
physical unit normal and let $e_A^i$, with $A=1,2$, denote an
orthonormal basis tangent to the boundary. We define the normal and
tangential projections
\begin{equation}
a_s=s^i a_i,
\qquad
a_A=e_A^i a_i,
\end{equation}
and
\begin{equation}
D_s
\equiv
s^iD_i.
\end{equation}
We impose the Sommerfeld-like conditions
\begin{equation}
\left(
\frac{1}{\alpha}\partial_\perp
+
D_s
+
\frac{1}{r}
\right)
a_A
=
0,
\qquad
A=1,2,
\label{eq:proca_bc_transverse}
\end{equation}
for the two transverse vector components, together with
\begin{equation}
\left(
\frac{1}{\alpha}\partial_\perp
+
D_s
+
\frac{1}{r}
\right)
\left(
a_s-\Phi
\right)
=
0,
\label{eq:proca_bc_longitudinal}
\end{equation}
for the longitudinal physical mode, and
\begin{equation}
\left(
\frac{1}{\alpha}\partial_\perp
+
D_s
+
\frac{1}{r}
\right)
Z
=
0
\label{eq:proca_bc_cleaning}
\end{equation}
for the incoming constraint-cleaning mode. 
Here $r$ denotes the coordinate distance from the asymptotic origin.
Using the evolution equation for $Z$, Eq.~(\ref{eq:Z_cleaning}), the last condition can equivalently be written as
\begin{equation}
\mathcal{G}
-
\kappa Z
+
D_s Z
+
\frac{Z}{r}
=
0.
\label{eq:proca_bc_gauss}
\end{equation}

These conditions are imposed only on the incoming characteristic degrees of freedom; the outgoing characteristic information is inherited from the bulk evolution. 
In particular, we do not impose independent component-wise Sommerfeld conditions on all eight evolved Proca variables.

The above prescription is a leading-order far-field approximation.
Because a massive Proca field has frequency-dependent asymptotic propagation, it is not an exact absorbing boundary condition, particularly for modes with frequencies close to the mass threshold.
The outer boundary is therefore placed sufficiently far from the binary that boundary effects do not influence the spacetime region and evolution times considered here.

\section{Sanity checks}
\label{sec:sanity_checks}

\begin{figure*}[htp!]
    \centering
    \includegraphics[width=\linewidth]{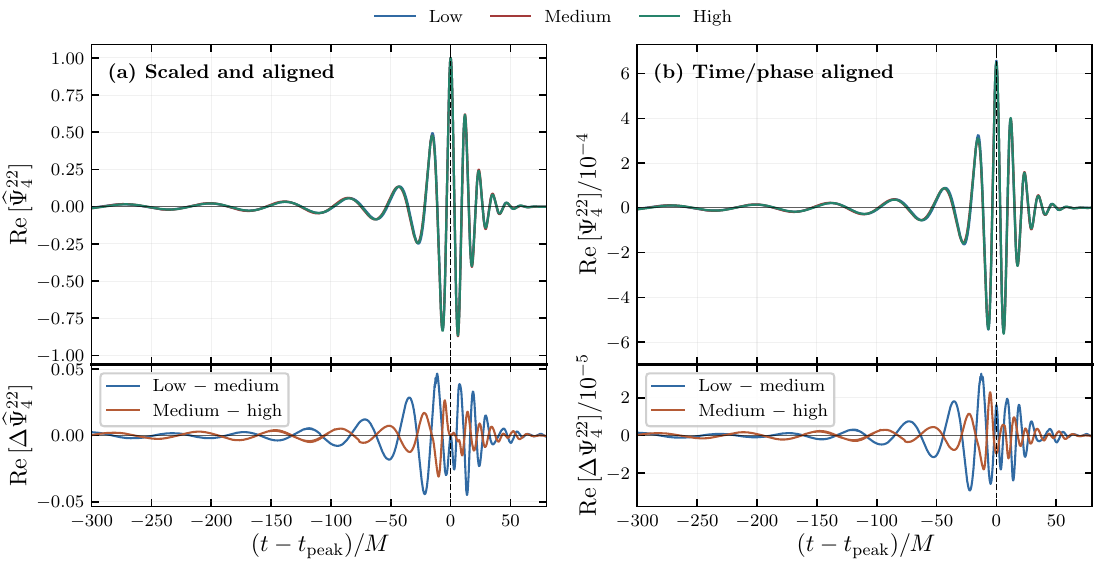}
    \caption{
    Convergence of the $(\ell,m)=(2,2)$ gravitational waveform for the $m_V=0.2$, $E_0=5\times10^{-4}M^{-1}$ simulation, extracted at $r=110M$. The waveforms are aligned such that the peak occurs at $t-t_{\rm peak}=0$ and their phases agree at the peak. (a) Each waveform is additionally normalized by its peak amplitude, isolating differences in waveform morphology. (b) The original amplitudes are retained. The lower panels show the differences between adjacent resolutions.
    }
    \label{fig:convergence_test}
\end{figure*}

\subsection{Convergence checks}

We assess numerical convergence using the $m_V=0.2$ configuration, for which simulations at low, medium, and high resolutions are available. The finest-grid spacings are $M/38.4$, $M/48$, and $M/60$, respectively. Table~\ref{tab:convergence} summarizes the waveform peak times and amplitudes at $r=110M$. Increasing the resolution systematically delays the waveform peak, from $t_{\rm peak}=1909.22M$ at low resolution to $2085.60M$ at high resolution. A similar delay with increasing resolution has been observed in long binary-black-hole numerical-relativity evolutions and is associated with accumulated numerical phase error~\cite{Husa:2007hp}.
Thus, to assess waveform convergence, we align the waveforms in time at their peak amplitudes and apply a constant phase rotation such that $\Psi_4$ is real at the peak, following the waveform-comparison procedures in \cite{Scheel:2008rj, Radia:2021smk}. 
This alignment removes the overall time and phase offsets, corresponding to extrinsic parameters over which waveform matches are maximized in gravitational-wave data analysis.

\begin{table}[t]
\caption{Resolution dependence of the $m_V=0.2$, $E_0=5\times10^{-4}M^{-1}$ simulation. The waveform quantities are evaluated from $\Psi_4^{22}$ extracted at $r=110M$.}
\label{tab:convergence}
\begin{ruledtabular}
\begin{tabular}{cccc}
Resolution & $\Delta x_{\rm min}/M$ & $t_{\rm peak}/M$ & $\max|\Psi_4^{22}|$ \\
\hline
Low    & $1/38.4$ & $1909.21875$ & $6.56860\times10^{-4}$ \\
Medium & $1/48$   & $2036.18750$ & $6.40705\times10^{-4}$ \\
High   & $1/60$   & $2085.60000$ & $6.50289\times10^{-4}$ \\
\end{tabular}
\end{ruledtabular}
\end{table}

Figure~\ref{fig:convergence_test} compares the $(\ell,m)=(2,2)$ waveforms over $-300M\leq t-t_{\rm peak}\leq80M$. 
For the scaled and aligned waveforms, we define
\begin{equation}
\widehat{\Psi}_{4,h}^{22}(\tau)
=
\frac{
\Psi_{4,h}^{22}\left(\tau+t_{{\rm peak},h}\right)
}{
\Psi_{4,h}^{22}\left(t_{{\rm peak},h}\right)
},
\qquad
\tau=t-t_{{\rm peak},h},
\end{equation}
where $h$ labels the numerical resolution. 
We quantify the difference between two resolutions $A$ and $B$ using the complex root-mean-square difference
\begin{equation}
\Delta_{\rm RMS}^{AB}
=
\left[
\frac{1}{\tau_2-\tau_1}
\int_{\tau_1}^{\tau_2}
\left|
\widehat{\Psi}_{4,A}^{22}(\tau)
-
\widehat{\Psi}_{4,B}^{22}(\tau)
\right|^2
\,d\tau
\right]^{1/2},
\label{eq:rms_waveform_difference}
\end{equation}
where $\tau_1=-300M$ and $\tau_2=80M$.
In the left panel, the waveforms are aligned in peak time and phase and normalized by their respective peak amplitudes. 
The complex root-mean-square differences decrease from $1.660\%$ between the low and medium resolutions to $0.819\%$ between the medium and high resolutions. 
In the right panel, only the peak time and phase are aligned, while the original amplitudes are retained. 
The corresponding root-mean-square differences decrease from $1.078\times10^{-5}$ to $5.624\times10^{-6}$. 
In both comparisons, the medium--high difference is approximately half the low--medium difference, indicating convergence with increasing resolution.

\subsection{Constraint violations}

\begin{figure*}[t]
    \centering
    \includegraphics[width=\textwidth]{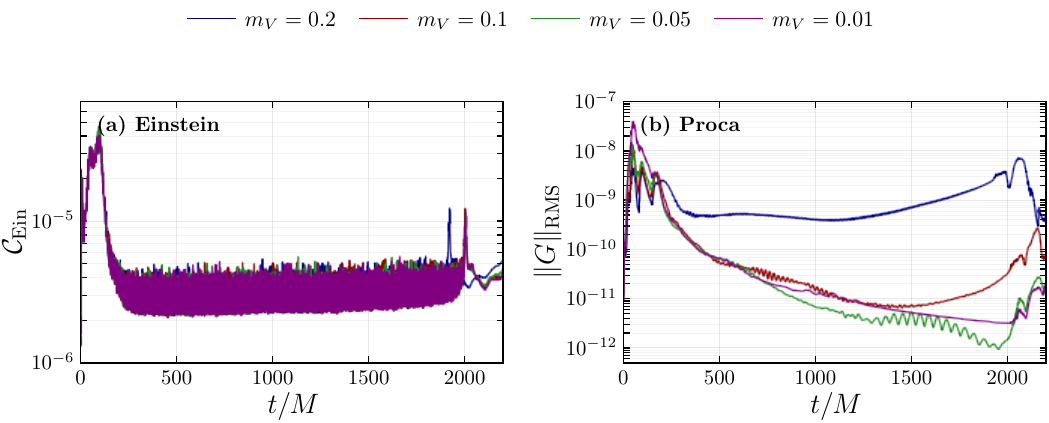}
    \caption{
    Evolution of constraint violations in the Einstein--Proca
    binary-black-hole simulations for different Proca masses $m_V$,
    with the initial electric-field amplitude fixed at
    $E_0=5\times10^{-4}$.
    The left panel shows the combined Einstein constraint norm,
    $\mathcal{C}_{\mathrm{Ein}}
    =\sqrt{\|\mathcal{H}\|_{\mathrm{RMS}}^2+
    \|\mathcal{M}\|_{\mathrm{RMS}}^2}$,
    where $\mathcal{H}$ and $\mathcal{M}$ denote the Hamiltonian
    and momentum constraints, respectively.
    The right panel shows the root-mean-square norm of the
    Proca constraint, $\|\mathcal{G}\|_{\mathrm{RMS}}$.
    The Einstein constraint norms are computed over unexcised
    grid nodes with equal weighting, whereas the Proca Gauss
    constraint is evaluated over the core-exterior region.
    }
    \label{fig:Einstein_Proca_Constraints}
\end{figure*}

Figure~\ref{fig:Einstein_Proca_Constraints} shows the evolution of the Einstein and Proca-field constraint violations for different vector masses. 
We quantify the Einstein constraint violation by the combined root-mean-square norm
\begin{equation}
\mathcal{C}_{\rm Ein}
=
\sqrt{
\left\|\mathcal{H}\right\|_{\rm RMS}^{2}
+
\left\|\mathcal{M}\right\|_{\rm RMS}^{2}
},
\end{equation}
where $\mathcal{H}$ and $\mathcal{M}_i$ are the Hamiltonian and momentum constraints, respectively. 
Their root-mean-square norms are defined by
\begin{equation}
\begin{split}
\left\|\mathcal{H}\right\|_{\rm RMS}
& =
\left[
\frac{1}{N}
\sum_{A=1}^{N}
\mathcal{H}_A^2
\right]^{1/2},\\
\left\|\mathcal{M}\right\|_{\rm RMS}
& =
\left[
\frac{1}{N}
\sum_{A=1}^{N}
\gamma^{ij}_A
(\mathcal{M}_i)_A
(\mathcal{M}_j)_A
\right]^{1/2},
\end{split}
\end{equation}
where $A$ labels the grid nodes included in the diagnostic and $N$ is their total number. 
By construction, $\mathcal{C}_{\rm Ein}$ is greater than or equal to either $\left\|\mathcal{H}\right\|_{\rm RMS}$ or the momentum-constraint norm $\left\|\mathcal{M}\right\|_{\rm RMS}$.
The Proca-field constraint $\mathcal{G}$ is defined by Eq.~\eqref{eq:Proca_constraint} in the main text, and its violation is quantified analogously by the root-mean-square norm $\|\mathcal{G}\|_{\rm RMS}$.
We monitor $\mathcal{C}_{\rm Ein}$ and $\|\mathcal{G}\|_{\rm RMS}$ over the numerical domain.
Following an initial transient, the Einstein constraint norm decreases to a few times $10^{-6}$ and remains at a comparable level throughout the inspiral, with localized increases near merger.
The Proca-field constraint exhibits a stronger dependence on the vector mass.
For $m_V=0.1$, $0.05$, and $0.01$, the constraint violations decrease to approximately $10^{-11}$ or below during the inspiral, whereas the $m_V=0.2$ case maintains a larger violation, of order $10^{-10}$--$10^{-9}$, which increases towards merger.
Overall, the constraint violations remain bounded throughout the numerical evolutions, with no sustained growth in the Einstein constraint norm.

\subsection{Light-crossing time checks}

We examine the arrival times of the peak of $|\Psi_4^{22}|$ at different extraction radii to check the propagation of gravitational waves in our simulations. We estimate the radial coordinate propagation speed between successive extraction radii as
\begin{equation}
v_j=\frac{r_j-r_{j-1}}{t_j-t_{j-1}},
\end{equation}
where $t_j$ is the coordinate time at which the peak of $|\Psi_4^{22}|$ reaches the extraction radius $r_j$.

\begin{figure}[t]
    \centering
    \includegraphics[width=\linewidth]{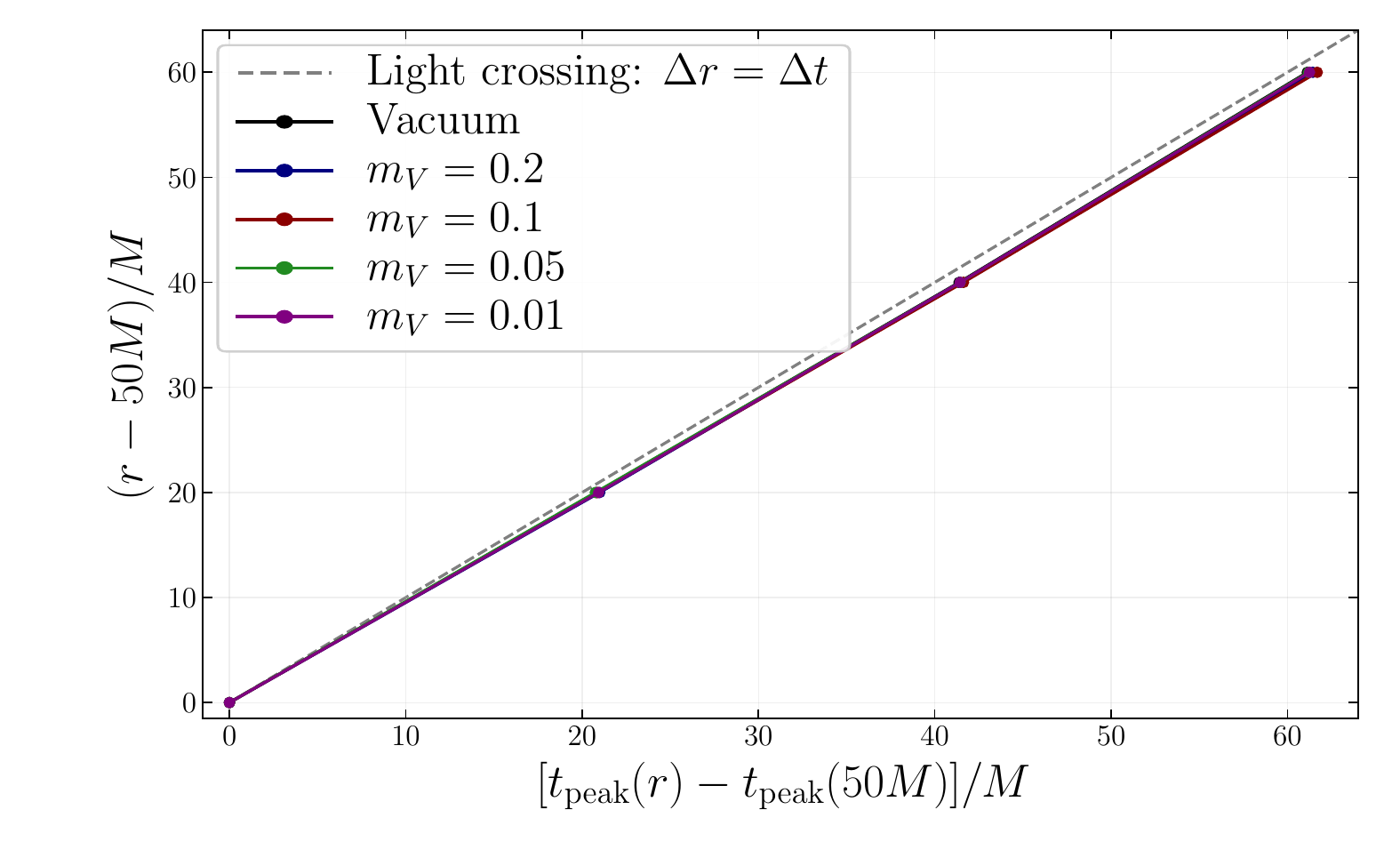}
    \caption{Difference in extraction radius, $\Delta r=r-50M$, as a function of the difference in the peak arrival time of $|\Psi_4^{22}|$, $\Delta t=t_{\rm peak}(r)-t_{\rm peak}(50M)$, for the vacuum and Proca-field simulations. The dashed line represents the light-crossing relation $\Delta r=\Delta t$.
    }
    \label{fig:Light_crossing_time}
\end{figure}

Figure~\ref{fig:Light_crossing_time} shows the differences in peak arrival times against the corresponding differences in extraction radii, with $r=50M$ as the reference. We find that the arrival-time differences approximately follow the light-crossing relation $\Delta r=\Delta t$, with inferred coordinate propagation speeds slightly below unity. Such deviations can arise from the spacetime curvature, the choice of coordinates, and finite-radius extraction effects. The results are consistent with the expected propagation of gravitational waves across the extraction radii.

\subsection{Symmetry-preservation check}

\begin{figure}[t]
    \centering
    \includegraphics[width=\linewidth]{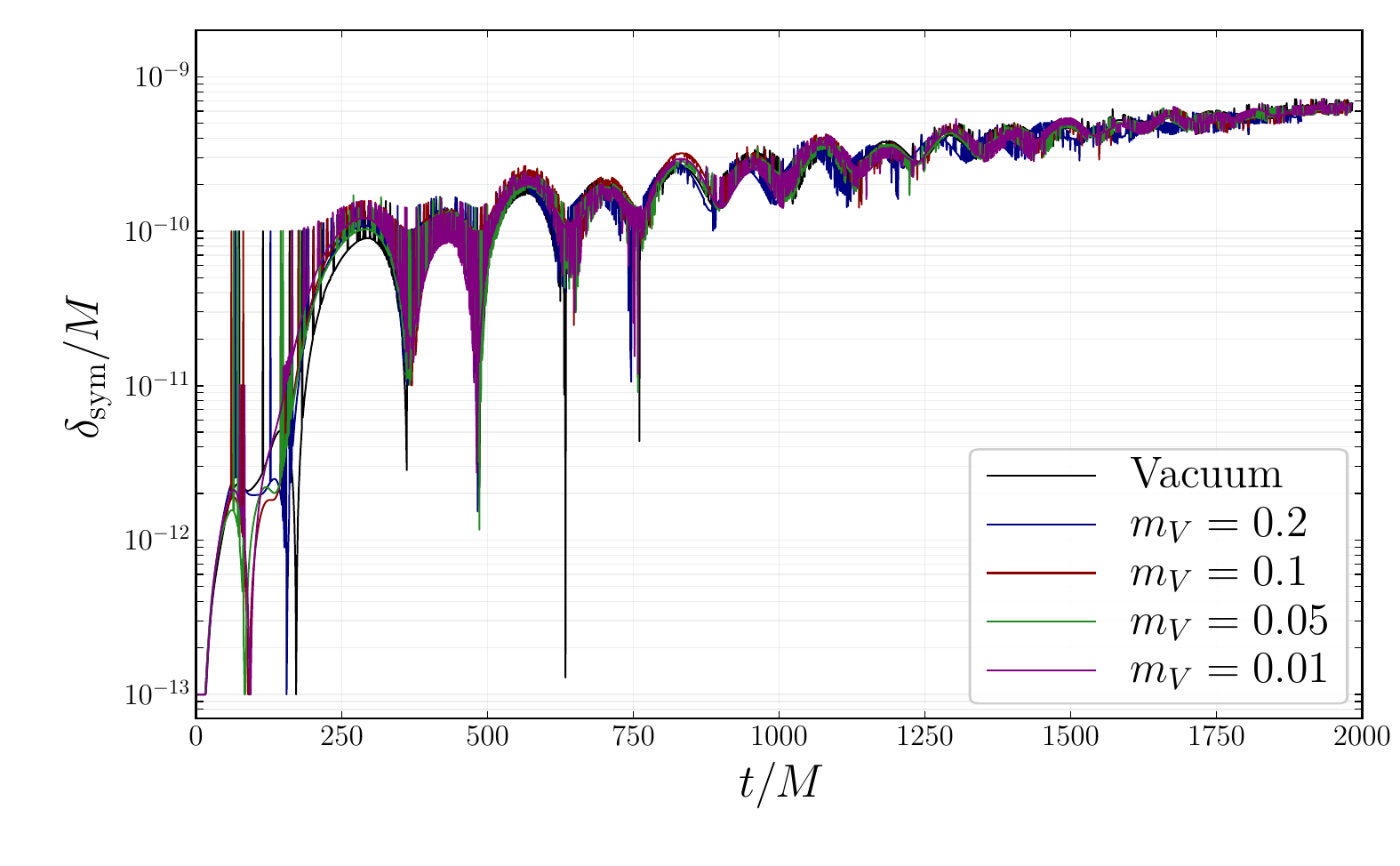}
    \caption{
    $\delta_{\rm sym}$, the coordinate distance between one black hole and the position of the other after a rotation by $\pi$ about the $z$-axis, as a function of time.
    }
    \label{fig:Reflection}
\end{figure}

Since the binary black holes have equal masses and vanishing spins, and the initial configuration is invariant under a rotation by $\pi$ about the $z$-axis, this symmetry should be preserved throughout the evolution. 
However, as mentioned in the main text, no symmetry is imposed on the simulations. 
Examining the preservation of this symmetry therefore provides a sanity check on the numerical evolution. 
We define the deviation from $\pi$-rotational symmetry about the $z$-axis,
\begin{equation}
\delta_{\rm sym} = \sqrt{(x_1+x_2)^2+(y_1+y_2)^2+(z_1-z_2)^2}.
\end{equation}
If the simulated binary preserves the rotational symmetry exactly, $\delta_{\rm sym}=0$. A nonzero value of \(\delta_{\rm sym}\) measures the violation of this symmetry in the numerical evolution.

Figure~\ref{fig:Reflection} shows the evolution of $\delta_{\rm sym}$, in units of $M$, during the numerical simulations. We find that $\delta_{\rm sym}\lesssim 10^{-9}M$ throughout the interval examined, approximately nine orders of magnitude smaller than the characteristic length scale $M$. This indicates that the $\pi$-rotational symmetry is preserved to high precision in our simulations.



\end{document}